\documentclass[%
 superscriptaddress,
 twocolumn,
 amsmath,amssymb,
 aps,
 prb,
floatfix,
]{revtex4-1}

\usepackage{epsfig}
\usepackage{amsmath,amssymb}
\usepackage[dvipsnames,usenames]{color}
\usepackage[normalem]{ulem}
\usepackage{tabularx}
\usepackage{siunitx}
\newcolumntype{Y}{>{\centering\arraybackslash}X}
\usepackage{soul} 
\usepackage{float}
\emergencystretch=\maxdimen
\usepackage{graphicx}

\usepackage{dcolumn}
\usepackage{bm}

\begin{document}


\title{Hybridization effect on the X-ray absorption spectra for actinide materials: Application to PuB$_4$}

\author{Wei-ting Chiu}
\email{wtchiu@ucdavis.edu}
\affiliation{%
Department of Physics, University of California, Davis, California 95616, USA
}%
\affiliation{
Theoretical Division, Los Alamos National Laboratory, Los Alamos, New Mexico 87545, USA}%

\author{Roxanne M. Tutchton}
\affiliation{
Theoretical Division, Los Alamos National Laboratory, Los Alamos, New Mexico 87545, USA}%

\author{Giacomo Resta}%
\affiliation{%
 Department of Physics, University of California, Davis, California 95616, USA
}%

\author{Tsung-Han Lee}
\affiliation{%
Physics and Astronomy Department, Rutgers University, Piscataway, New Jersey 08854, USA}

\author{Eric D.~Bauer}
\affiliation{Materials Physics and Application Division, Los Alamos National Laboratory, Los Alamos, New Mexico 87545, USA}

\author{Filip Ronning}
\affiliation{Institute for Materials Science, Los Alamos National Laboratory, Los Alamos, New Mexico 87545, USA}

\author{Richard T. Scalettar}%
\affiliation{%
 Department of Physics, University of California, Davis, California 95616, USA
}%


\author{Jian-Xin Zhu}
 \email{jxzhu@lanl.gov}
\affiliation{
Theoretical Division, Los Alamos National Laboratory, Los Alamos, New Mexico 87545, USA}%
\affiliation{
Center for Integrated Nanotechnologies, Los Alamos National Laboratory, Los Alamos, New Mexico 87545, USA}%


\date{\today}

\begin{abstract}
Studying the local moment and 5$f$-electron occupations sheds insight into the electronic behavior in actinide materials. X-ray absorption spectroscopy (XAS) has been a powerful tool to reveal the valence electronic structure when assisted with theoretical calculations. However, the analysis currently taken in the community on the branching ratio of the XAS spectra generally does not account for the hybridization effects between local $f$-orbitals and conduction states. In this paper, we discuss an approach which employs the DFT+Gutzwiller rotationally-invariant slave boson (DFT+GRISB) method to obtain a local Hamiltonian for the single-impurity Anderson model (SIAM), and calculates the XAS spectra by the exact diagonalization (ED) method. A customized numerical routine was implemented for the ED XAS part of the calculation. By applying this technique to the recently discovered  5$f$-electron topological  Kondo insulator PuB$_4$, we determined the signature of 5$f$-electronic correlation effects in the theoretical X-ray spectra. We found that the Pu 5$f$-6$d$ hybridization effect provides an extra channel to mix the $j=5/2$ and $7/2$ orbitals in the 5$f$ valence.  As a consequence,  the resulting electron occupation number and spin-orbit coupling strength deviate from the intermediate coupling  regime.
\end{abstract}

\maketitle


\section{Introduction}

The actinide solids are essential materials for nuclear power generation. In addition to their important energy applications, the valence electronic structure of the 5$f$-electron series exhibits a range of complex and fascinating emergent behavior of fundamental physical interest. In the early actinide elements the 5$f$-electrons are itinerant while the valence electrons in the late actinides are more localized. Among the 5$f$ series, Pu is the most complex solid which has six allotropic crystal phases at ambient pressure. Furthermore, it undergoes a 25\% volume expansion from its $\alpha$ phase to $\delta$ phase. These exotic phenomena of Pu have puzzled condensed matter experimentalists and theorists for decades. Different theoretical and modeling approaches have been used to unravel the physics of this complicated elemental solid.~\cite{Weinberger1985, Savrasov01, Terry02, Dai03, Kotliar_DMFT_RMP, Sci.Rep.5.15958,Adv.phy.68.1} On top of that, Pu-based compounds also show exotic behaviors such as lack of temperature dependence (up to 1000 K) of the magnetic susceptibility in PuO$_2$ solid, which can only be explained with conspiring spin-orbit coupling, crystal field splitting, and electronic correlation effects.~\cite{2015Kolorenc,Gendron_2017} In addition, the interplay between spin-orbit coupling and correlation effects from the localized $f$-electrons makes the $f$-electron compounds natural candidates for topological Kondo insulators, for which SmB$_6$ and PuB$_6$ are good examples.~\cite{Lu2013, PhysRevB.97.201114}  In particular, Pu-based compounds stand out compared to their 4$f$ counterparts due to a larger energy scale in the spin orbit coupling, giving rise to measurable bulk gaps.~\cite{PhysRevLett.104.106408, Dzero16, Shim12, Booth10205, PhysRevLett.111.176404, Zhang1464}

Studying the electronic structure is essential towards understanding the basic physical properties of solids, such as magnetism, superconductivity, 
and topology,~\cite{RevModPhys.81.235} and the covalent interactions in molecules.\cite{Sergentu_2018, Ganguly_2020} In order to probe the properties of the valence electrons, X-ray absorption spectroscopy (XAS) has been demonstrated to be a powerful tool for this purpose.\cite{PhysRevB.76.073105, RevModPhys.81.235, Sergentu_2018, Ganguly_2020} Previous analyses on the branching ratio of the XAS spectra mostly assumed a single $f^n$ configuration for the valence $f$ states and used spin-orbit sum-rules to estimate the ratio.~\cite{PhysRevB.53.14458, PhysRevB.57.112} Recently, it has been suggested~\cite{Shim12,Booth10205} that multiconfigurational $f$-orbital states could play an important role in the electronic structure of U- and Pu-based material systems. It is therefore  important to gain  a complete picture of the valence electron behavior. In this work, we focus on analyzing the Pu-$N_{4,5}$ absorption edges. The theoretical XAS spectra are calculated from a model derived using the first-principle DFT+GRISB method and that account for interactions between electrons, which compete with the hybridization between 5$f$-orbital and conduction electrons.  This approach enables for an in depth analysis of the multiconfigurational nature of the 5$f$ states.

\begin{figure}
\includegraphics[width=\linewidth]{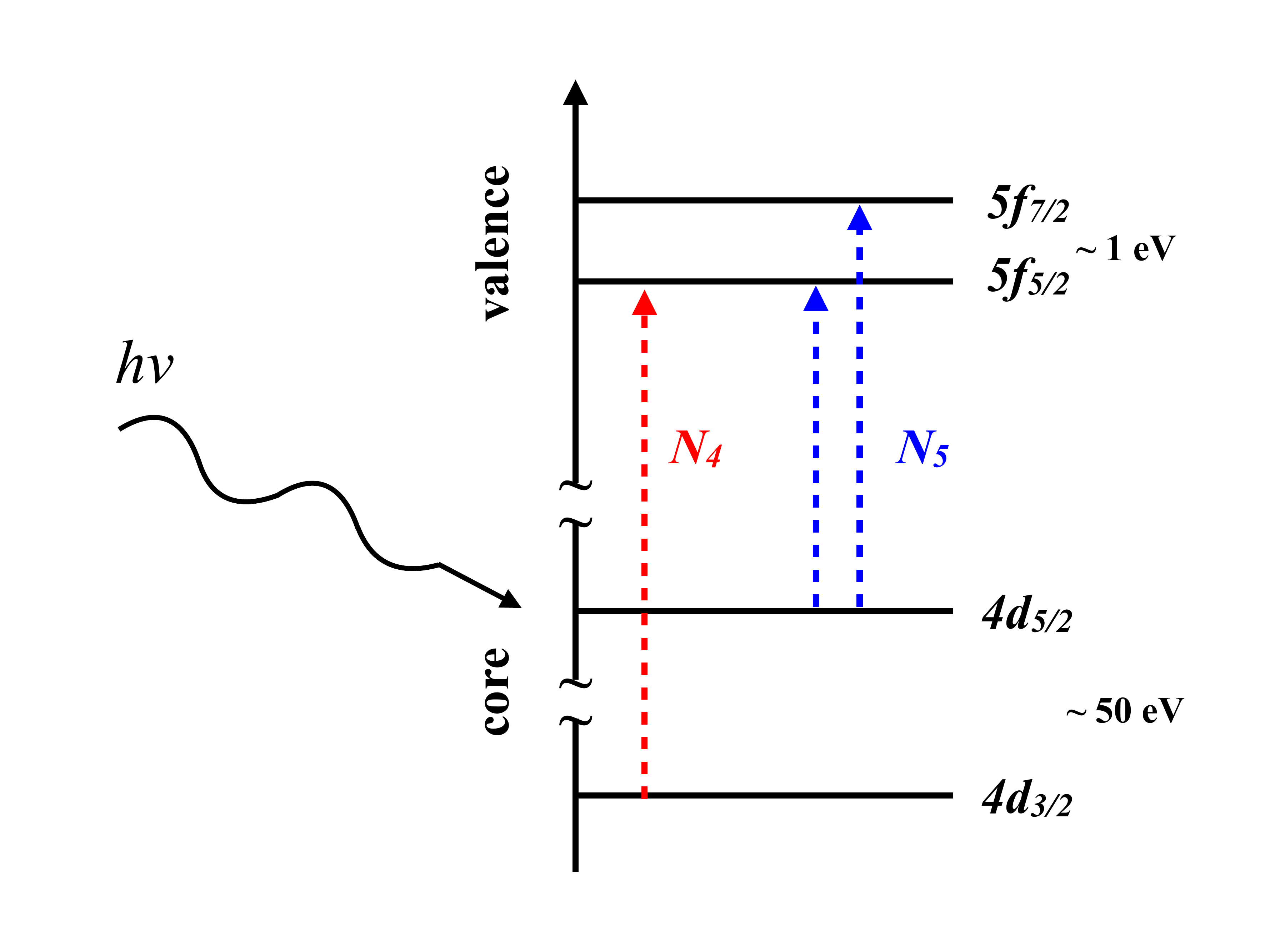}
\caption{Schematic diagram for $4d \rightarrow 5f$ XAS process. A photon excites a $4d$ core electron to a $5f$ valence level. The $N_{4,5}$ absorption edge contains two peaks separated by $\sim$ 50 eV which can be clearly resolved in the spectra.}
\label{fig:x-ray-scheme}
\end{figure}

For the actinide $N_{4,5}$ $(4d \rightarrow 5f)$ absorption edge, the $N_4$ $(d_{3/2})$ and $N_5$ $(d_{5/2})$ peaks are clearly distinct since the 4$d$-orbital core-hole has a much stronger  spin-orbit coupling than the electrostatic interaction. We can obtain the intensities, $I(N_4)$ and $I(N_5)$, by fitting the peaks with Lorentzian functions without too much interference. Due to the dipole selection rule, $d_{3/2}$ core electrons can only be excited to an empty $f_{5/2}$ level while $d_{5/2}$ electrons are allowed to transit to both $f_{5/2}$ and $f_{7/2}$ levels. A schematic diagram for the $N_{4,5}$ edge is shown in Fig.~\ref{fig:x-ray-scheme}. Thus, the $N_{4,5}$ branching ratio, $B=I(N_5)/[I(N_5)+I(N_4)]$, serves as a good probe for the 5$f$ valence occupations. Previous studies~\cite{RevModPhys.81.235, PhysRevB.53.14458, PhysRevLett.93.097401, Shim_2009} have applied the spin-orbit sum rule on the $N_{4,5}$ branching ratio to determine the 5$f$ electron fillings on the angular momentum levels. However, the sum-rule analysis approach assumed an integer 5$f$ valence filling, which implies that the electrons are fully localized, and the atomic multiplet calculations consider only an integer total number of valence electrons per atom.~\cite{PhysRevB.53.14458, PhysRevB.57.112, RevModPhys.81.235, PhysRevB.43.13401, PhysRevB.32.5107}  A more complete description of the 5$f$ electronic structure requires taking into account the hybridization of 5$f$-electrons with conduction orbitals and other valence electrons. However, the Hilbert space for the XAS calculation grows exponentially as the number of orbitals increases. If certain itinerant orbitals are explicitly considered as bath orbitals that can hybridize with the localized orbitals,  then they need to be directly included in the XAS Hamiltonian, making the calculation computationally prohibitive. Moreover, the number of bath orbitals required to reach convergence remains an open question that needs to be examined carefully.~\cite{2015Kolorenc}


Recently the density functional theory  plus Gutzwiller rotationally invariant slave boson (DFT+GRISB) method  has proved to be a reliable and efficient tool for describing the strong electronic correlation effects in heavy fermion systems, giving quantitatively accurate results when compared to more computationally expensive techniques such as dynamical mean-field theory (DMFT).~\cite{PhysRevLett.111.196801,PhysRevX.5.011008,PhysRevLett.118.126401}The advantage of this method is that it  maps the original lattice problem onto an effective two-site embedding impurity Hamiltonian, consisting of a physical and an auxiliary bath orbital with proper hybridization, through the process of Schmidt decomposition.~\citep{PhysRevX.5.011008,PhysRevB.96.235139,PhysRevB.99.115129} In this paper, we develop a numerical approach for calculating the XAS spectra within the  DFT+GRISB framework, in which the embedding two-site (i.e., one local site and one bath site)
impurity Hamiltonian is used as a basis for constructing the model for the XAS calculation, insuring that hybridization effects from the surrounding atomic environment are included. The central ingredient of this technique is as follows: We start with the embedding Hamiltonian obtained from the DFT+GRISB ground state calculation, and then introduce a core-hole Hamiltonian for the final state. Then, we apply the dipole transition operator for the $d\rightarrow f$ transition  and use a Lanczos algorithm\cite{lanczos_Meyer} to calculate the spectral function. A custom numerical implementation utilizing an exact diagonalization (ED) method was developed for the XAS part of the calculation. By applying this technique to PuB$_4$, we show that   
this costly calculation can be carried out on present midscale computing cluster within a reasonable amount of time.

The outline of the remaining text of this paper is as follows: In Sec.~\ref{sec:theory-method}, we give a detailed step by step procedure for setting up the effective Anderson impurity Hamiltonian for the XAS spectroscopy within the DFT+GRISB framework.  In Sec.~\ref{sec:results}, the electronic band structure and X-ray absorption spectral density, together with their sensitivity to the Pu-$5f$ electron valence, are presented.  In addition, the effect of the hybridization of Pu-5$f$ electron correlated orbitals with the itinerant valence orbitals is also discussed. A summary of our main findings are given in Sec.~\ref{sec:conclusion}.

\begin{figure}[h]
\begin{center}
\includegraphics[width=\linewidth]{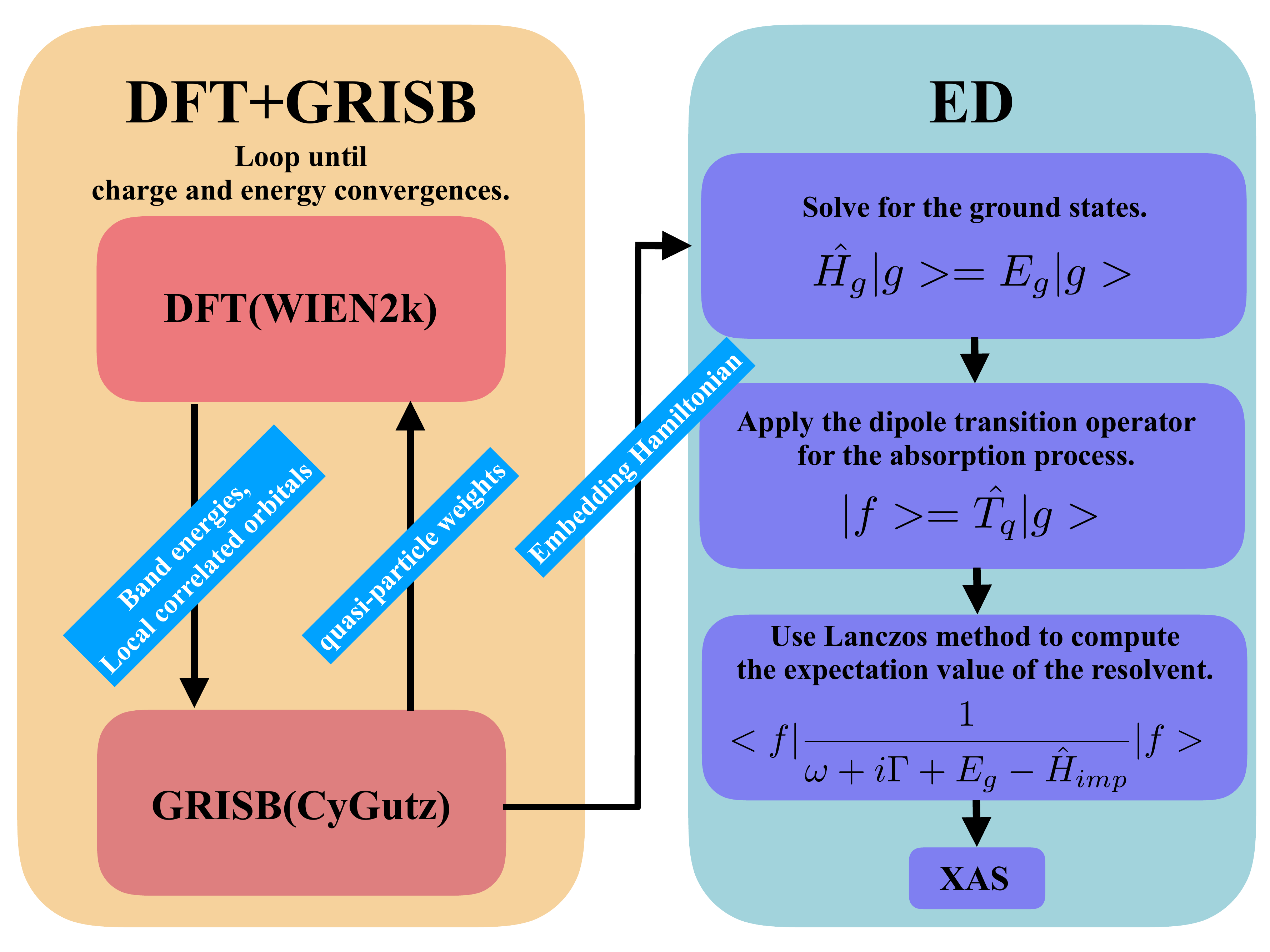}
\end{center}
\caption{Overview of our numerical procedure for calculating XAS which consists of two major parts: (1) WIEN2k and CyGutz cycle is run until charge and energy convergence is reached. The calculation gives the correlated band structure, density of states, and the local embedding Hamiltonian. (2) The embedding Hamiltonian is fed into an ED calculation for the XAS. First, the ground state is calculated from the embedding Hamiltonian. Then the dipole transition operator is applied to get the final state. Finally, the Lanczos method is applied to compute the spectra. }
\label{fig:cygutz-ed}
\end{figure}

\section{Theoretical Method}
\label{sec:theory-method}

\subsection{DFT+GRISB}

We calculated the band structures of PuB$_4$ using DFT+GRISB as implemented in the CyGutz package.~\cite{PhysRevX.5.011008,PhysRevLett.118.126401} The method utilizes the full-potential linear augmented plane-wave basis  WIEN2K code~\citep{wien2k} for the DFT band structure calculation. We use the exchange-correlation functional in the PBE generalized gradient approximation.~\cite{PhysRevLett.77.3865} Throughout the work, all calculations on PuB$_4$ were performed with experimentally determined lattice parameters and atomic positions.~\cite{PhysRevB.97.201114}  After the convergence of DFT, the many-body Hamiltonian for the strongly correlated $f$-orbital is then constructed by projecting the Kohn-Sham Hamiltonian to the correlated orbitals.~\citep{PhysRevB.81.195107,PhysRevX.5.011008} The many-body Hamiltonian is a generic multi-orbital Hubbard model,

\begin{equation}
\hat{H}=\sum_{\mathbf{k}i,j,a,b}\varepsilon_{\mathbf{k}ij}^{ab} \hat{c}_{\mathbf{k}ia}^{\dagger}\hat{c}_{\mathbf{k}jb}+\sum_{\color{black}i}H_{\text{loc}}[\{\hat{c}_{ia},\hat{c}_{ia}^{\dagger}\}],\label{eq:Hubbard_kspace}
\end{equation}
\noindent where $a$ and $b$ are the spin and orbital indices,  which contain both the correlated $f$-orbitals and the other non-correlated orbitals, and $i$ and $j$ are the indices labeling the atoms within the unit cell.  The generic Hubbard Hamiltonian is then solved using the GRISB algorithm.~\citep{PhysRevX.5.011008,PhysRevLett.118.126401} 

Within  the GRISB framework, the original lattice many-body problem is mapped onto two auxiliary Hamiltonians. The first one is the quasiparticle Hamiltonian,
\begin{equation}
\hat{H}_{qp} = \sum_{\mathbf{k}} \Big[ R_{ia\alpha}\varepsilon_{\mathbf{k}ij}^{\alpha\beta}R^{\dagger}_{j\beta b}+\lambda_{iab} \delta_{ij} \Big] \hat{Q}_{\mathbf{k}i a}^\dagger \hat{Q}_{\mathbf{k}j b}\;, 
\label{eq:Hqp}
\end{equation}
\noindent where $\hat{Q}^{\dagger}$  ($\hat{Q}$) are quasiparticle creation (annihilation) operators,  $R$ corresponds to the quasiparticle renormalization matrix, $Z=R^\dagger R$, and $\lambda$ is the renormalized potential. The $R$ and $\lambda$ matrices have a block diagonal structure in the atom index $i$, whose respective block is labeled by $R_{i}$ and $\lambda_{i}$.~\cite{PhysRevX.5.011008} Here the repeated indices of $i$, $j$, $a$, $b$, $\alpha$, and $\beta$ imply summation. The other is an auxiliary embedding  impurity Hamiltonian,
\begin{eqnarray}
&& \hat{H}_{i,\text{emb}}= \hat{H}_{\text{loc}}\big[\{\hat{c}_{i\alpha}^{\dagger},\hat{c}_{i\alpha}\}\big] 
 \nonumber\\
&&\qquad +\sum_{\alpha a}\big(\mathcal{D}_{i_{a\alpha}}\hat{c}_{i\alpha}^{\dagger}\hat{f}_{ia}+\text{H.c.}\big)+\sum_{ab}\lambda_{iab}^{c}\hat{f}_{ib}\hat{f}_{ia}^{\dagger},\label{eq:Hemb}
\end{eqnarray}
\noindent where $c_{i}$ is the physical fermionic operators on the impurity and $f_{i}$ is the fermionic operator for the bath for atom $i$. The matrix $\mathcal{D}_{i}$ describes the hybridization between the bath and impurity orbitals, and $\lambda^c_{i}$ corresponds to the energy level for the bath. The mapping from the lattice problem to the auxiliary impurity problem can be realized through a Schmidt decomposition. The auxiliary impurity problem consists of a collection of a physical atom and a bath representing the local  environments.~\citep{PhysRevX.5.011008,PhysRevB.96.235139,PhysRevB.99.115129}  It can be solved efficiently using the exact diagonalization technique. 
The parameters $R$, $\lambda_c$, $\mathcal{D}_{i}$, and $\lambda^c_i$ for these two auxiliary Hamiltonians are determined self-consistently by solving the following saddle-point equations,~\cite{PhysRevLett.118.126401}
\begin{eqnarray}
&&\hskip-1.0em\Delta_{i,ab}^{p}=\frac{1}{\mathcal{N}}\sum_{\mathbf{k}}\Pi_{i} \big[ f_T(R\varepsilon_{\mathbf{k}}R^{\dagger}+\lambda)\big]_{ba} \Pi_{i},\label{eq:SB_SP_eq1}\\
&&\hskip-1.4em\big[\Delta^{p}_{i}(1-\Delta^{p}_{i})\big]^{1/2}_{ac}\mathcal{D}_{i,c\alpha} \nonumber \\
&&=\frac{1}{\mathcal{N}} \big[ R_{i}^{-1} \sum_{\mathbf{k}}\Pi_{i}R\varepsilon_{\mathbf{k}}R^{\dagger}f_T(R\varepsilon_{\mathbf{k}}R^{\dagger}+\lambda) \Pi_{i}\big]_{\alpha a},\label{eq:SB_SP_eq2}\\
&&\hskip-1.0em\sum_{cb\alpha} \frac{\partial}{\partial d^p_{is}}\big[\Delta^p_{i}(1-\Delta^p_{i})\big]^{\frac{1}{2}}_{cb}\big[\mathcal{D}_i\big]_{b\alpha}\big[R_i\big]_{c\alpha}+\text{c.c} \nonumber \\
&&\qquad\qquad\qquad\qquad\qquad\hskip3.1em +\big[l+l^c\big]_{is}=0,\label{eq:SB_SP_eq3}\\
&&\hskip-1.0em\hat{H}_{i,\text{emb}}|\Phi_{i}\rangle=E^{c}_{i}|\Phi_{i}\rangle,\label{eq:SB_SP_eq4}\\
&&\hskip-1.4em\Big[\mathcal{F}^{(1)}_{i}\Big]_{ab}\equiv\langle\Phi_{i}|\hat{f}_{ib}\hat{f}_{ia}^{\dagger}|\Phi_{i}\rangle-\Delta_{i,ab}^{p}=0,\label{eq:SB_SP_eq5}\\
&&\hskip-1.4em\Big[\mathcal{F}^{(2)}_{i}\Big]_{\alpha a}\equiv\langle\Phi_i|\hat{c}_{i\alpha}^{\dagger}\hat{f}_{ia}|\Phi_i\rangle- R_{ic\alpha}\big[\Delta^{p}_i(1-\Delta^{p}_i)\big]_{ca}^{\frac{1}{2}}\nonumber \\ 
&&\hskip18.1em =0.\label{eq:SB_SP_eq6}
\end{eqnarray}
\noindent The symbol $f_{T}$ stands for the Fermi function of a single-particle matrix at temperature $T$, the $\Pi_{i}$ means the projection onto the atom $i$.  We utilized the following matrix parameterizations,

\begin{align}
\Delta^p_{i}&=\sum_s d^p_{is}\, h_s^{\text{Transpose}},\,\,\,\lambda^c_{i}=\sum_s l^c_{is} h_s,\nonumber \\ 
\lambda_{i}&=\sum_s l_{is} h_s,\,\,\,R_{i}=\sum_s r_{is} h_s\label{eq:R_param},
\end{align}
\noindent where the set of matrices $h_s$ are an orthonormal basis of the space of Hermitian matrices (with respect to the canonical trace inner product). The parameters $d^p_{is}$, $l^c_{is}$ and $l_{is}$ are real, while $r_{is}$ is complex. The RISB saddle-point equations can be viewed as a root problem of finding $R$ and $\lambda$ that satisfy Eqs. (\ref{eq:SB_SP_eq5}) and (\ref{eq:SB_SP_eq6}), which can be solved using any root finding algorithms.

After the convergence of $R$ and $\lambda$, the GRISB charge density is computed and embedded back to the DFT loop. The charge self-consistent DFT+GRISB loop is shown schematically in the left of Fig.~\ref{fig:cygutz-ed}. We set the self-consistency condition for the total energy error to less than $10^{-6}$ and charge error to less than $10^{-3}$. We employ the standard fully localized limit form for the double-counting functional.\cite{Anisimov_1997_2} For the ground-state calculations, we use the values for $U_f$=4.5 eV and $J_f$=0.5 eV and consider the fully rotational invariant Coulomb interaction in the local Hamiltonian for Pu-5$f$ electrons.

\subsection{Impurity Hamiltonian}
We calculate the XAS from the single-impurity Anderson Hamiltonian (SIAM),~\cite{AmentRMP}
\begin{equation}
    \hat{H}_{SIAM}=\hat{H}_{emb}+\hat{H}_{core-hole}.
\end{equation}
It contains two parts. The first piece, $\hat{H}_{emb}$, is exactly the embedding Hamiltonian for the local impurity from the DFT+GRISB result, while the second term, $\hat{H}_{core-hole}$, expresses the interactions between the local valence electrons and the core-hole created in the XAS process. 


The core-hole Hamiltonian is written as
\begin{equation}
\begin{split}
    \hat{H}_{core-hole}
    &=\sum_u \epsilon_u \hat{d}^{\dagger}_u \hat{d}_u + \sum_{ijkl}U^{ccdd}_{ikjl}\hat{c}^{\dagger}_i\hat{d}^{\dagger}_j\hat{d}_l\hat{c}_k\\
    &+\sum_{ijkl}U^{cddc}_{ikjl}\hat{c}^{\dagger}_i\hat{d}^{\dagger}_j\hat{c}_l\hat{d}_k,
\end{split}
\end{equation}
where $\epsilon_{u}$ is the on-site energy for the core-hole orbitals, $d^{(\dagger)}$ is the annihilation (creation) operator for the core-hole, $U^{ccdd}$ ($U^{cddc}$) is the Coulomb (exchange) interaction matrix between correlated and core orbitals, and i, j, k, l are combined indices for the orbital and spin states.

\subsection{Dipole Transition Matrix}

In the XAS process when an X-ray  is incident upon a material with its energy tuned to a certain absorption edge,  a core electron can be excited to a valence orbital leaving a core-hole behind. This process follows Fermi's golden rule. The electric dipole transition for q-polarized light is given by~\cite{AmentRMP, cowan, PhysRevB.53.14458, RevModPhys.81.235}
\begin{equation}
    \hat{T}_q=\sum_{m_v,m_c}\langle l_vs;j_vm_v|\bold{r}_q|l_cs;j_cm_c\rangle\hat{c}^{\dagger}_{j_v,m_v}\hat{d}_{j_c,m_c},
\end{equation}
where
\begin{equation}
    \begin{split}&\langle l_vs;j_vm_v|\bold{r}_q|l_cs;j_cm_c\rangle=(-1)^{j_c-m_c+l_c+s+j_v+1}\\
&\times[j_c, j_v]^{1/2}\begin{Bmatrix}
 j_c & 1 & j_v \\
l_v & s & l_c \end{Bmatrix}
\begin{pmatrix}
 j_c & 1 & j_v \\
m_c & q & -m_v \end{pmatrix}
\langle l_vs||\bold{r}||l_cs\rangle\;.
\end{split}
\end{equation}
In the above equations, $c^{\dagger}_{j_v, m_v}$ indicates the creation operator of a correlated valence electron with total angular momentum $j_v$ and spin angular momentum $m_v$, $d_{j_c, m_c}$ is the annihilation operator for a core-hole at total angular momentum $j_c$ and spin angular momentum $m_c$, $l_v(c)$ is the orbital angular momentum for the valence (core) orbital, $s=\frac{1}{2}$  is the spin of the electrons, and $[a,b,...]$ is a shorthand of $(2a+1)\times(2b+1)\times...$, and  $\{\}$and () stand for Wigner-9j and Wigner-6j expressions, respectively. We remark that there is a difference in the square-root pre-factor on the right-hand side of Eq. (14) compared to that in the literature.~\cite{RevModPhys.81.235, PhysRevB.53.14458, PhysRevLett.93.097401} We derive Eq. (14) by combining Eqs. (14.28) and (14.47) from the book by Cowan,~\cite{cowan} and only with the pre-factor derived in this work could we obtain the X-ray absorption spectra that satisfy the spin-orbit sum-rule shown in Table I.

\subsection{XAS Intensity}


\begin{figure*}
\includegraphics[width=\linewidth]{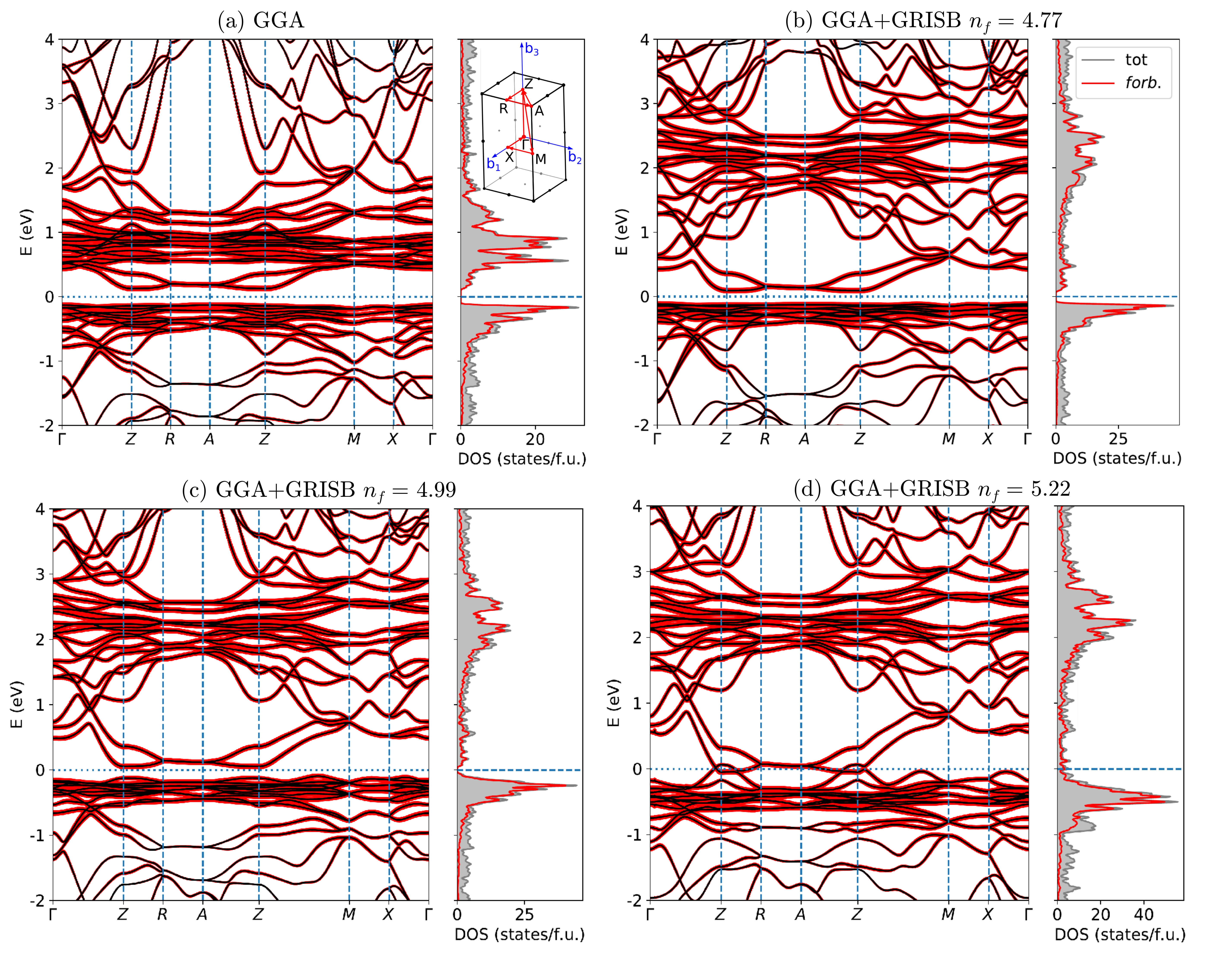}
\caption{Band structures for PuB$_4$. (a) GGA results. The number of 5$f$ valence electron is $n_f \sim 4.7$. (b) GGA+GRISB results with $n_f=4.77$. (c) GGA+GRISB results with $n_f=4.99$. (d) GGA+GRISB results with $n_f=5.22$. Inset: Path along high symmetry points in momentum space.}
\label{fig:bandstructure}
\end{figure*}

\begin{figure}
\centering
\includegraphics[width=0.8\linewidth]{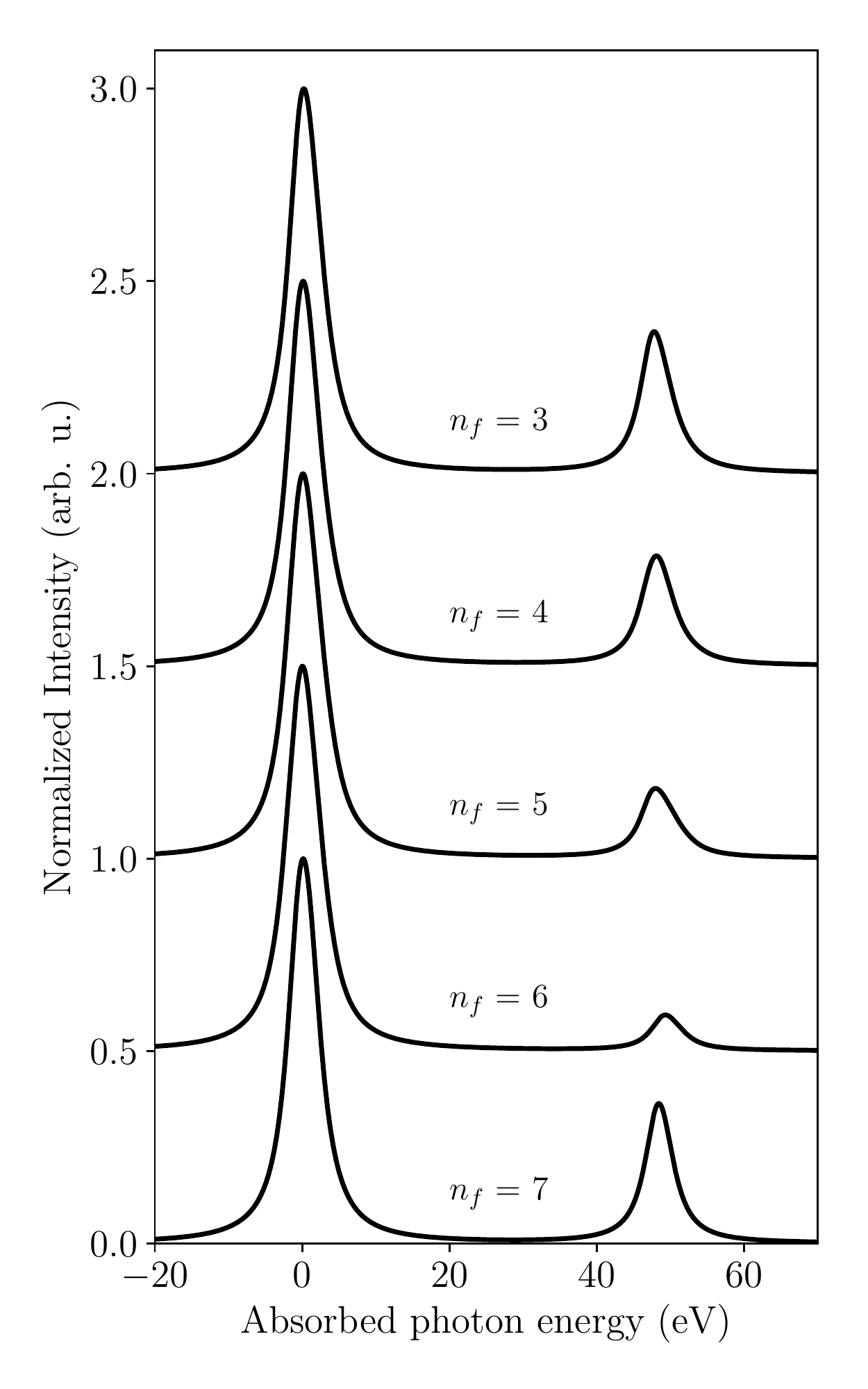}
\caption{Theoretical XAS from different valence occupations calculated using the ED method without hybridization. In the calculation, we use values for Pu$^{3+}$ taken from Cowan's code for parameters such as Slater integrals and spin-orbit constant. The results are similar to the atomic multiplet calculations. The spectra for $n_f=3 \sim 6$ are shifted vertically for clearer comparison.}
\label{fig:xas-1}
\end{figure}

\begin{figure}
\includegraphics[width=0.8\linewidth]{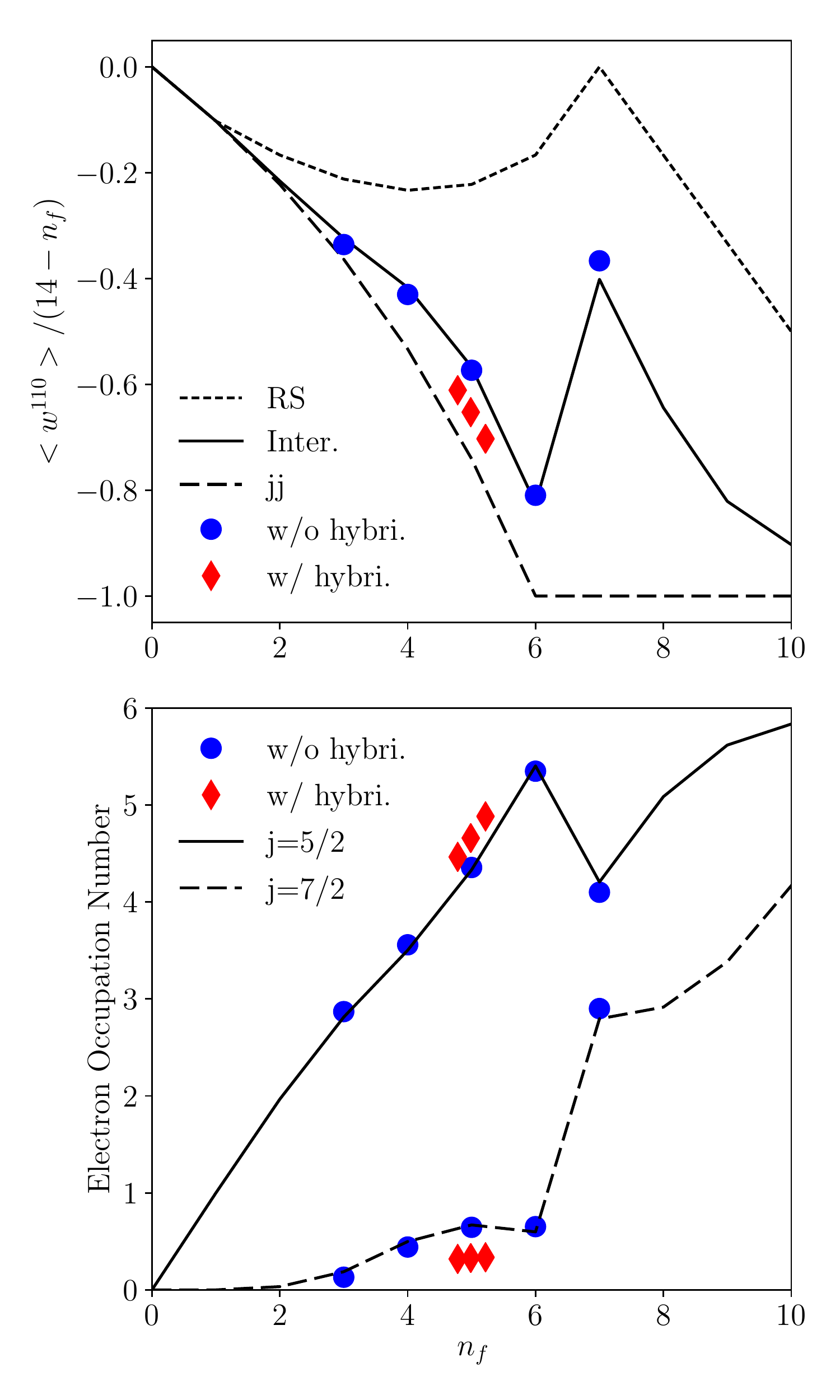}
\caption{Upper panel: Spin-orbit coupling strength per hole. The lines depict the theoretical values in Russel-Sanders (RS), intermediate (Inter.), and $jj$-coupling schemes(jj). The values for the dots are derived from the calculated XAS branching ratio using the spin-orbit sum rule. Lower panel: Electron occupation numbers on $j=5/2$ and $j=7/2$ orbitals. The lines are results from intermediate coupling. The dots are from the spin-orbit sum rule with corresponding total number of valence electrons.}
\label{fig:soc-nf}
\end{figure}

\begin{figure}
\includegraphics[width=\linewidth]{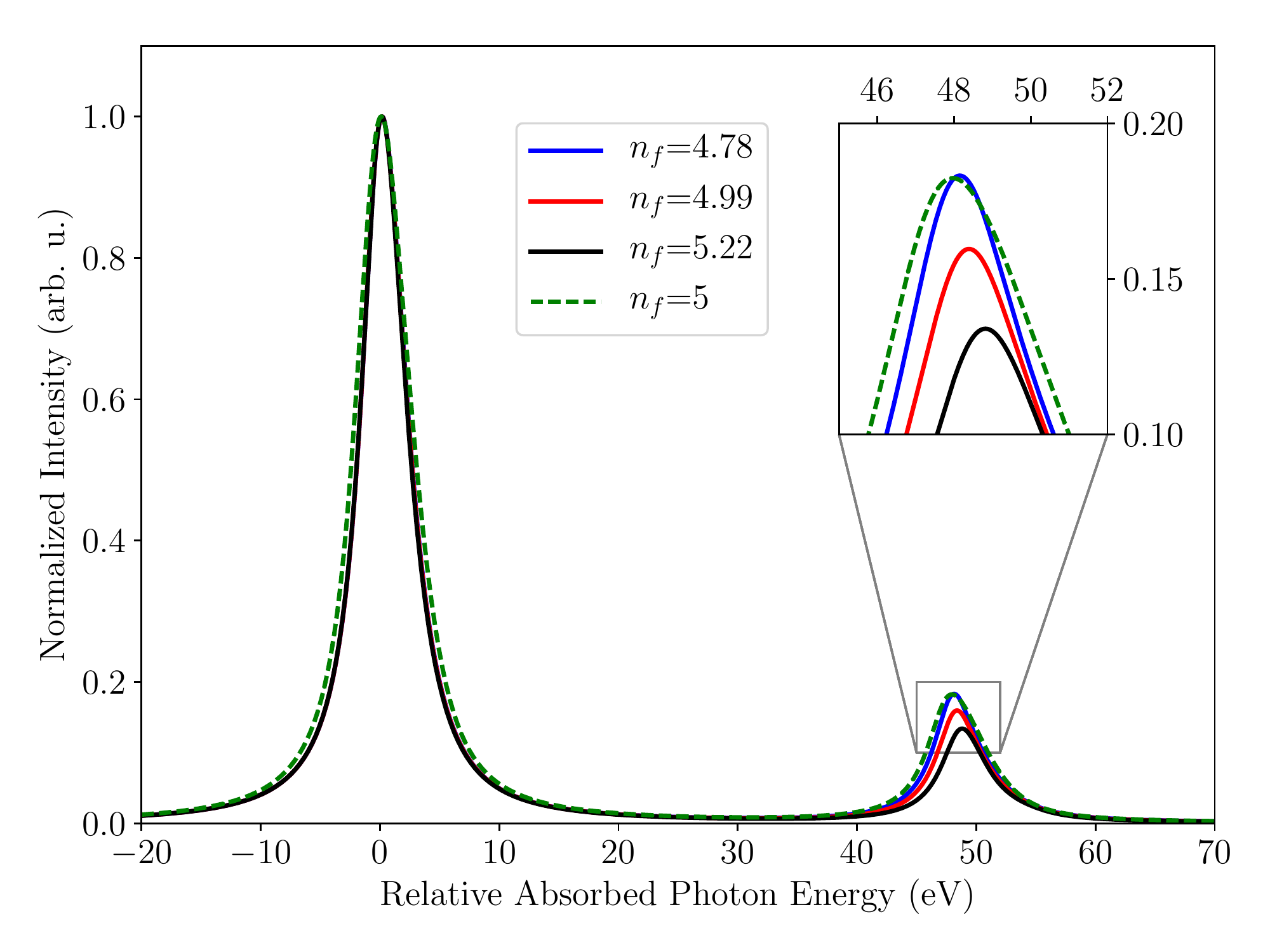}
\caption{XAS for different valence occupations. The $n_f=5$ spectrum (dashed line) is calculated using the single-configuration atomic multiplet model and is provided for comparison. The  spectra for $n_f=4.77$, 4.99, and 5.22 (solid lines) are calculated using a multi-configuration Hilbert space. The intensities are normalized to the maximum peak value.}
\label{fig:xas}
\end{figure}

With the dipole transition operator, we can calculate the XAS intensity by\cite{AmentRMP, lanczos_Meyer}
\begin{equation}
    I(\omega)\sim\frac{1}{\pi}\sum_q \text{Im}[\langle g|\hat{T}^{\dagger}_q\frac{1}{\omega+i\Gamma+E_g-\hat{H}_{SIAM}}\hat{T}_q|g\rangle],
\end{equation}
where $\omega$ is the photon frequency and $\Gamma$ is the Lorentzian broadening. The ground state and its corresponding energy are obtained by
\begin{equation}
    \hat{H}_g|g\rangle=\hat{H}_{emb}|g>=E_g|g\rangle.
\end{equation}

\subsection{Computation Procedure}

A schematic overview of our approach for calculating XAS is shown in Fig.~\ref{fig:x-ray-scheme}. We first obtain the band structures by running the WIEN2k and CyGutz cycle until both charge and energy convergence is reached. Then, we take the embedding Hamiltonian from the DFT+GRISB solution as the local Hamiltonian for the single impurity in our system. With the local Hamiltonian we could build the single impurity Anderson Hamiltonian (SIAM) and solve for the ground state by the exact diagonalization method. After that, the dipole transition operator is applied on the ground state to simulate the effect of an X-ray incident on the sample as in experiments. A core electron absorbs the X-ray and is promoted to a valence orbital, and the system transitions to a higher energy final state. The X-ray absorption spectra can be obtained by computing the expectation value of the resolvent for the single-particle spectral function using the Lanczos method.

\section{Results and Discussions}
\label{sec:results}

While there are no photoemission spectroscopy for band structure and density of states, as well as XAS measurements for the Pu valence in PuB$_4$ yet, we calculated these physical properties using the DFT+GRISB approach with varying 5$f$ occupations. This flexibility of the approach enables to track directly the sensitivity of these properties to the Pu-5$f$ occupancy.  It also lends the future experimental measurements an opportunity to directly compare with the beyond-DFT theory to determine the valence electronic states for PuB$_4$.

Figure~\ref{fig:bandstructure}(a) shows the DFT band structure. The bare DFT result is the same as in Ref.~\onlinecite{PhysRevB.97.201114} and reproduced here. The number of electrons in the 5$f$ orbital is 4.7 and the bulk band gap is 254 meV. Fig.~\ref{fig:bandstructure}(b), (c), and (d) are the DFT+GRISB results with calculated $n_f$ = 4.77, 4.99, and 5.22, respectively. The inset in Fig.~\ref{fig:bandstructure}(b) illustrates the $k$-path used for plotting the band structures in the first Brillouin zone. The introduction of correlation effects using the DFT+GRISB method results in three major changes in the band structure as compared to the bare DFT result. Similar changes were likewise observed in previous DFT+GRISB band structure calculations for SmB$_6$~\cite{Lu2013} (i) The $j$=7/2 bands move  from around 1 eV above the Fermi energy (due to the spin-orbit coupling) in the DFT calculation to $\sim$ 2.0 eV in the DFT+GRISB results. As a result, the band structure around the Fermi energy is dominated by 5$f_{5/2}$ and 6$d$ orbitals. (ii) The bandwidth of the $j$=5/2 bands near the Fermi level in the DFT+GRISB calculation are reduced slightly compared to the DFT results. The quasi-particle renormalization weight in the $j$=5/2 sector is around $Z\sim$0.79, which is consistent with the reduction of the $j$=5/2 bandwidth. (iii) The band gap is reduced following the  introduction of electronic correlations. The size of the band gap can be further reduced by increasing the number of 5$f$ electrons in the DFT+GRISB calculations and can even be closed by filling the 5$f$ $j$=5/2 band above $n_f$=5. At $n_f=5$, the band gap is renormalized to 138 meV which is closer to the transport measurement of 35 meV~\cite{PhysRevB.97.201114} than 254 meV gap in the bare DFT calculation.
The aforementioned correlation effects are the consequence of the GRISB approximation where the R matrix renormalize the bandwidth of the correlated band and the $\lambda$ shifted the position of the correlated bands. 

For calculating non-hybridized theoretical X-ray spectra, $D$ and $\lambda^c$ were set to zero.  The parameters for the local impurity's spin-orbit coupling constant, core-hole  spin-orbit coupling constant, and the Slater integrals for the valence-core Coulomb interactions ($U^{ccdd}_{ikjl}$ and $U^{cddc}_{ikjl}$) in $\hat{H}_{core-hole}$ were taken from Cowan's relativistic Hartree-Fock code.~\cite{cowan} We used values from the computed result of Pu$^{3+}$ and the Coulomb interactions are decreased by 20\%. The direct core-hole potential ($U^{ccdd}_{iijj}$) is set to the same value as $U_f$. We broadened the spectra by 2 eV to take into account the core-hole 
life-time effect.~\cite{PhysRevB.35.2667} However, the computational cost for calculating the XAS spectrum for the multi-$f^n$ electron configurations increases dramatically as compared to the single electron configuration case because the size of the Hilbert space for the ground state grows from a few thousand in single-configuration case to tens of billions for the multi-configuration case with $n_f$ ranging from 3 to 7. The situation is worse for getting the final states with an extra core-hole. In order to obtain results in a reasonable amount of time, we can consider only the direct Coulomb term for the core-valence interaction. With this simplification, the full Hamiltonian becomes block diagonal. The contribution to the spectral function by each block can then be calculated independently and summed together to obtain the final result. This simplification does not have noticeable effect on the branching ratio since it only caused a minor energy level splitting (roughly 0.1 $\sim$ 0.5 eV) within each peak of the $N_{4,5}$ edge, and these extra lines were buried in the 2 eV broadening. 

The calculated non-hybridized X-ray absorption spectra  of Pu $5f^3$ to $5f^7$are shown in Fig.~\ref{fig:xas-1}.  We consider various valence fillings and obtained each absorption spectrum separately. The results are qualitatively consistent with the spectra calculated previously on $\alpha$-Pu.~\cite{RevModPhys.81.235, PhysRevB.53.14458} 
When the XAS is calculated from the ground state with single $f^n$ configuration, that is, no hybridization between local 5$f$ orbitals and the other orbitals is considered, the spin-orbit sum-rule can be applied to find the electron occupation number for both the $j=5/2$ and $j=7/2$ levels. The sum-rule~\cite{RevModPhys.81.235, PhysRevB.53.14458} is given by
\begin{equation}
\frac{\langle W^{110}\rangle}{14-n_f} - \Delta = -\frac{5}{2}\left( B-\frac{3}{5} \right),
\end{equation}
where $\Delta$ is a small correction term for the $N_{4,5}$ edges (usually within 3\%)\cite{PhysRevLett.93.097401} and we set to zero for convenience, the quantity $B$ is the branching ratio, which are obtained by integrating the intensity of the two Lorentzian peaks, while the expectation value of the tensor operator for the spin-orbit coupling is given by~\cite{PhysRevB.53.14458}
\begin{equation}
\label{eq:w110}
\langle W^{110} \rangle = \frac{2}{3}\langle l \cdot s\rangle = n_{7/2} - \frac{4}{3}n_{5/2}\;.
\end{equation}
Using the sum-rule and the fact that
\begin{equation}
n_f = n_{7/2} + n_{5/2}\;,
\end{equation}
the occupations in $f_{5/2}$ and $f_{7/2}$ valence levels can be found. The determined values are listed in Table~\ref{tab:e-config}. We note here that although the XAS technique itself involves the change of 5$f$ electron count in the final state, the branching ratio analysis determines the ground state 5$f$ occupation of the system. We can also calculate the 5$f$ occupations from the ground state obtained by ED using
\begin{equation}
n_j = \sum_{i \in j} \langle g| \hat{f}^\dagger_i \hat{f}_i |g \rangle\;.
\end{equation}
The ground state electron occupation for the correlated orbitals as determined from the ED calculations is in excellent agreement with the spin-orbit sum-rule estimations. Fig.~\ref{fig:soc-nf} illustrates the spin-orbit coupling strength and the electron occupation numbers for Pu with varying $n_f$. For non-hybridized cases the results follow the intermediate coupling scheme for 5$f$ systems. There are minor discrepancies in the SOC strength per hole between our calculated and the intermediate coupling curve, which is due to the fact that we used the same SOC constant for Pu in all our calculations while the values of the curve were obtained from each element. Pu's non-hybridized result ($n_f$=5) lies on the intermediate coupling curve perfectly.

\begin{table}[]
\caption{Results from single $f^n$ configuration. $B$ is the branching ratio from Fig.~\ref{fig:xas-1}. The quantity $\langle W^{110}\rangle$ is calculated by Eq.~(\ref{eq:w110}). The Pu-$5f$ electron occupation numbers $n_{5/2}$ and $n_{7/2}$ are obtained by either the spin-orbit sum rule or from the ED ground state. The two sets of numbers agree well with each other.}
\label{tab:e-config}
\begin{tabularx}{0.95\linewidth}{>{\centering\arraybackslash}X  | >{\centering\arraybackslash}X  >{\centering\arraybackslash}X  | >{\centering\arraybackslash}X  >{\centering\arraybackslash}X  | >{\centering\arraybackslash}X  >{\centering\arraybackslash}X  l}
\toprule
 &  &  & \multicolumn{2}{c}{Sum rule}  & \multicolumn{2}{c}{ED ground state} \\
$n_f$  &  $B$ & $\bigl< W^{110} \bigr>$ & $n_{5/2}$  & $n_{7/2}$  & $n_{5/2}$  & $n_{7/2}$   \\
\hline
3 & 0.73 &  -3.69 & 2.87 & 0.13 & 2.84 & 0.16\\
4 & 0.77 &  -4.30 & 3.56 & 0.44 & 3.54 & 0.46\\
5 & 0.83 &  -5.16 & 4.36 & 0.65 & 4.35 & 0.65\\
6 & 0.92 &  -6.48 & 5.35 & 0.65 & 5.32 & 0.68\\
7 & 0.75 &  -2.57 & 4.10 & 2.90 & 4.08 & 2.92\\
\toprule
\end{tabularx}
\end{table}

\begin{table*}[]
\caption{Weights in each occupation from multi-configuration ground state obtained from ED calculations for PuB$_4$ when the hybridization is taken into account. The orbital dependent Pu-$5f$ electron occupation from the sum rule is also listed. *Result for $\delta$-Pu.}
\label{tab:e-config-hyb}
\begin{tabularx}{0.9\linewidth}{ Y | Y Y Y Y Y | Y Y | Y Y }
\toprule
$n_f$&\multicolumn{5}{c}{Weight} & \multicolumn{2}{c}{ED} & \multicolumn{2}{c}{Sum rule}    \\
total & 3 & 4& 5 & 6 & 7 & $n_{5/2}$ & $n_{7/2}$  & $n_{5/2}$ & $n_{7/2}$ \\
\hline
4.78 & 0.03 & 0.30 & 0.54 & 0.13 & 0.01 & 4.45 & 0.33 & 4.46 & 0.32 \\
4.99 & 0.01 & 0.21 & 0.57 & 0.20 & 0.01 & 4.65 & 0.34 & 4.66 & 0.33\\
5.22 & 0.01 & 0.13 & 0.53 & 0.32 & 0.02 & 4.87 & 0.35 & 4.88 & 0.34\\
\hline
5.18* & 0.01 & 0.13 & 0.56 & 0.29 & 0.02 & 4.79 & 0.38 & 4.80 & 0.37\\
\toprule
\end{tabularx}
\end{table*}

Finally, we calculated the XAS for PuB$_4$ with hybridization between local 5$f$ orbitals and the bath orbitals. The embedding Hamiltonian was extracted from the  DFT+GRISB calculations. Fig. ~\ref{fig:xas} shows the calculated XAS for the hybridized cases with various values of Pu occupation. As is evident, the spectral density for the $N_5$ edge depends on $n_f$. The spectrum for the hybridized case of $n_f$=4.99 has a larger branching ratio of $B=0.86$ compared to the non-hybridized $n_f$=5 case with  $B$ = 0.83. The reason for this difference is that the hybridization provide an extra channel to mix the $j=5/2$ and $j=7/2$ valence spin-orbitals. As a result, more spectra weight is transferred from the higher $j=7/2$ to the lower $j=5/2$ level, and thus the $N_4$ intensity is smaller for the hybridized case with the same number of valence electrons. The calculated $n_{5/2}$ and $n_{7/2}$ values are shown in Fig.~\ref{fig:soc-nf} and Table~\ref{tab:e-config-hyb}.  In the presence of hybridization, about 0.3 electron is transferred to the $j=5/2$ levels for $n_f$ $\sim$ 5. The calculated weight from the multi-configuration ED ground state is shown in Table~\ref{tab:e-config-hyb}. More than half of the weights are in the $f^5$ configuration. With the  increase in the average Pu-5$f$ electron occupancy, the configuration weights are mostly transferred from $f^4$ to $f^6$. There is a slight difference in the values for $n_f$ between the GRISB ground state method and the ED ground state. The reason for this is that in the GRISB calculations, we used $f^2$ to $f^7$ electron configurations for $n_f$ = 4.77,  $f^3$ to $f^7$ for $n_f$ = 4.99, and $f^3$ to $f^8$ for $n_f$ = 5.22. However, for the ED calculations, we considered only $f^2$ to $f^7$ electron configurations for all three cases with the $n_f$ value  shown in Table~\ref{tab:e-config-hyb}. There is a less than 1\% difference between the two approaches. We used  $f^3$ to $f^7$ configurations in the ED method for the XAS to dramatically reduce the computational cost. 

If the average Pu-5$f$ occupancy of the system is tuned from an integer valence filling of $n_f=5$, then most of the change in valence occupation results from the change in $j=5/2$ levels and the filling of $j=7/2$ stay stable. Thus, the intensity of $N_4$ is roughly linearly dependent on $n_{5/2}$, as shown in Fig.~\ref{fig:xas}. One might notice that the energy difference between the two peaks increases as $n_f$ grows. The $N_4$ peak shifts from $\sim $48 eV when $n_f$=4.78 ($B=0.84$) to $\sim$ 49 eV when $n_f$=5.22 ($B=0.88$). The shift can be explained by the core-valence Coulomb interaction.  The attractive Coulomb interaction between the core-hole and the $j=5/2$ levels is stronger when there was more valence electrons within these orbitals. As a result, the energy difference between $|\overline{4d}^1 5f^{n+1}, j=5/2>$ and $|\overline{4d}^1 5f^{n+1}, j=7/2>$ final states becomes larger, resulting in a larger splitting between the two peaks.
Finally, for this example of PuB$_4$, one can see from Fig.~\ref{fig:soc-nf} that when the valence-bath hybridization is taken into account,  the system is no longer in the intermediate coupling regime but in between intermediate coupling and $jj$-coupling schemes.

We have also applied the same approach to the $\delta$-Pu, the orbital-dependent Pu-$5f$ occupation from both ground state ED calculations and XAS sum rule, as well as the ground state electron configuration weight are included together with PuB$_4$ in Table~\ref{tab:e-config-hyb}. The $N_{4,5}$ XAS branch ratio $B=0.87$ (for $n_f=5.18$). As one can see the orbital-resolved Pu-$5f$ occupations from the ED and the sum rule reference matches. In addition, the obtained configuration weights from our ED method are in a reasonable agreement with the earlier LDA+DMFT calculations~\cite{Shim12,2007Zhu} and the experimental measurements on $\delta$-Pu,~\cite{ Booth10205} providing further evidence of a mixed-valent $5f$ state~\cite{JGTobin:2007} in $\delta$-Pu. 
 For PuB$_4$, a resonant X-ray emission spectroscopy (RXES) $L_3$-edge measurement  by Booth {\em et al.}  suggested a multi-configurational state of Pu-5$f$ with dominant  $f^4$ and smaller $f^5$ configuration fractions.  Comparing the theoretical XAS to the experimental results should enable us to extract the Pu-5$f$ electron occupancy. In our calculations of XAS spectroscopy, we do see a significant increase of $f^4$ weight fraction when the Pu-5$f$ occupancy is reduced away from $n_f=5$. However, the DFT based calculations cannot reproduce the experimental Pu-5$f$ occupancy in this compound, and therefore, further theoretical studies are required to resolve this issue.


\section{Conclusion}
\label{sec:conclusion}
In summary, we have developed a technique for computing the XAS beyond  the atomic multiplet model. The technique takes into account not only correlated orbitals but also their hybridization with the ligand orbitals from the DFT+GRISB solution. The DFT+GRISB  solution allowed us to analyze physical properties, such as band structure, density of states, and conductivity of materials with electronic correlations. Hybridization effects are important in strongly-correlated materials and it is therefore essential to include them in theoretical calculations of electronic response. In our results for PuB$_4$, we find that correlation effects renormalize the band gap to a size closer to the experimental measurement.~\cite{PhysRevB.97.201114} Furthermore, in the XAS calculations the hybridization between correlated and other conduction or valence orbitals provides another channel to mix the two angular momentum orbitals, which changes the $N_{4,5}$ branching ratio, shifting it from intermediate coupling towards $jj$-coupling. 
 This research opened a new avenue to calculate the X-ray spectroscopy including a complete description of electronic behavior.

\acknowledgments
We thank Corwin Booth, Jindrich Kolorenc, and Jean-Pierre Julien for helpful discussions. In particular, we are grateful to Corwin Booth for sharing with us the RXES observation on PuB$_4$ before publication.  One of us (J.-X.Z.) acknowledges the hospitality of CNRS through the CPTGA Program
at Universite de Grenoble Alpes during his stay as a visiting scientist. This work was carried out under the auspices of the U.S. Department of Energy
(DOE) National Nuclear Security Administration (NNSA) under Contract No. 89233218CNA000001.
It was supported by the G. T. Seaborg Institute (W.-T.C.) and the LANL LDRD Program (W.-T.C. \& R.M.T.),  the NNSA Advanced Simulation and Computing
Program (J.-X.Z.),  and in part supported by Center for Integrated Nanotechnologies, a DOE BES user facility, in partnership with LANL Institutional Computing Program for computational resource. F.R. and E.D.B. acknowledge support from the DOE BES “Quantum Fluctuations in Narrow Band Systems” project. 
T.H.L. was supported by the Department of Energy under Grant No. DE-FG02-99ER45761. G.R. was supported by NSF DMR Grant No. 1411336. The work of R.T.S. was supported by the grant DE-SC0014671 funded by the U.S. Department of Energy, Office of Science.

\bibliography{wtchiu_xas}

\end{document}